\documentstyle[12pt,titlepage,epsf]{article}
 \newcommand{\beq}{\begin{equation}}
 \newcommand{\eeq}{\end{equation}}
 \newcommand{\beqa}{\begin{eqnarray*}}
 \newcommand{\eeqa}{\end{eqnarray*}}

 \def\pabl#1#2{\frac{\partial {#1}}{\partial { #2}}}

\begin{document}
\begin{center}
{\Large The conformation of conducting polymer chains:\\ Hubbard polymers\\}
\vspace{5cm}
M.Otto and T.A.Vilgis\\[12pt]
Max-Planck-Institut f\"ur Polymerforschung\\ 
Postfach 3148, D-55021 Mainz, F.R.G.\\
\end{center}
\vspace{10ex}
\noindent
{\bf Abstract}:
 
\vspace{2ex}
\noindent
The conformational and electronic properties of conducting flexible random and
self-avoiding polymer chains are under investigation. A Hamiltonian for
conjugated flexible polymers is introduced and its physical consequences are
presented. One important result is that the electronic degrees of freedom
greatly affect the conformational statistics of the walks and vice versa. The
electronic degrees of freedom extend the size of the chain. The end-to-end
distance behaves as $R\propto L^{\nu}$ with $\nu=(d+1)/(d+2)$, where $d$ is
the spatial dimension.
 
\newpage
\setcounter{equation}{0}
\noindent
The properties of conducting flexible polymers are widely unknown. 
On the contrary, quite detailed models exist for linear
completely stretched (e.g. crystalline) polymers. In such cases it has been argued that the charge 
transfer happens only along the chain. 
 Most theoretical 
investigations on conducting or conjugated (both terms are used
interchangeably here) polymers have been devoted to
one-dimensional chains studying mainly conductive properties, 
especially since the SSH model by Su
et al. \cite{su:79}\cite{su:80}. This Hubbard-type theory predicts the existence 
of solitons which appear
as mobile domain walls between degenerate ground states of the 
dimerized chain. More recent work focusses on the influence of impurities 
and non-linear excitations \cite{harigaya:90}\cite{fesser:92}, and on the stability 
of polarons depending on chain ends and other effects that 
break the conjugation \cite{mizes:93}. Experiments with flexible conjugated polymer chains in solution (e.g. 
polydiacetylenes) indicate a
rod-coil transition at room temperatures \cite{harrah:85}\cite{lim:86a}. 
Moreover, the structural and conformational dependence of conductivity of these 
chains has 
been investigated (see e.g. \cite{wegner:89}). 
These experiments are in particular 
very interesting, as an important interplay between electronic properties and
the conformational properties is demonstrated. Obviously the theories cited
above cannot be applied to these cases, as the polymers are flexible, rather
than quasi-one-dimensional. 
 
As a first step in this direction a simple scaling 
argument has been put forward by Pincus et al. \cite{pincus:87}. Their idea is to relate this phenomenon to an 
interplay between the delocalized $\pi$-electron system and 
conformational entropy. A simple free energy Ansatz has been proposed that
uses a tight binding approximation for the one electron band together with a simple assumption for the chain entropy, i.e. it depends linearly on the segment length $l$ . By minimization of the free energy an optimum 
Kuhn segment length, $l^\ast$, is found, which behaves roughly as
$l^\ast (r)\simeq l_c\left(r+\frac{1}{2}\right)$. $r$ is the number of 
electrons on a given segment and 
$l_c=\left(\frac{2\pi^2 t}{3\alpha T}\right)^{\frac{1}{3}}$ where $t$ is 
the transfer integral, $\alpha$ a constant of proportionality and $T$ 
temperature.
 
Other theoretical work on the conformation of conducting polymer chains 
 concerns 
the stabilization of rod-like conformations by an interaction between the
delocalized electron structure and a polarizable solvent 
\cite{schweizer:86a} \cite{schweizer:86b}.
 
The purpose of this paper is to introduce and show the consequences of a Hamiltonian that contains the conformational properties of the polymer chains and the electronic information of the charge carriers. The chains are assumed to be flexible 
and thus excitations such as phonons play only a minor role, if e.g.
 a self-avoiding walk is assumed as a model for the bare chain 
with electrons present. 
The Hamiltonian presented in the following describes the conformation of 
conducting polymer chains as a result of chain entropy {\bf and} the presence of 
delocalized $\pi$-electrons along the chain contour. 
Our theory elaborates thus on the scaling results
by Pincus et al. \cite{pincus:87}. It is shown below that the suggested
rod-to-coil transition does not take place although the walk stretches
significantly. The Hamiltonian put forward here of a flexible chain containing electronic
degrees of freedom consists of two parts. The
first is the Wiener-Edwards Hamiltonian for a self-avoiding walk, the second
is a Hubbard type Hamiltonian that describes the electrons. The electrons are
confined to hop along the self-avoiding chain (contour hopping) in the
continuum limit. The latter assumption is not necessary, and can be easily relaxed to account for more general situations, where the charge transfer can also take place between 
excluded volume monomers for a single chain, forming current bridges
 (bridge hopping), or
between different chains (inter chain hopping). 
 
The starting point is a lattice model for a chain in a 
$d$-dimensional hypercubic lattice 
of lattice constant $a$. The electrons hop between lattice points which are 
nearest neighbors and are constrained to remain within the range $a$ of the
chain.
Neglecting electron spin and Coulomb correlations between electrons, the interaction between conformational and electronic
degrees of freedom is written as:
\beq
\label{ort}
\beta H_{el.}=-t\int_0^L ds \sum_{{\bf x}}\sum_{\mu=1}^d\left(
c^{\dagger}({\bf x})c({\bf x}+a{\bf e}_{\mu}) +
c^{\dagger}({\bf x})c({\bf x}-a{\bf e}_{\mu})+ h.c.\right)
\delta_{{\bf x},{\bf r}(s)}.
\eeq
$t$ is the transfer integral scaled by inverse temperature, $\beta$. It has 
the dimension of inverse length and is supposed to be invariant over the contour;
also $t>0$. Therefore we require $t=\beta t_0/L$ where $t_0$ is the transfer
integral of the SSH model in the case of no dimerization (i.e. $u_n=u_{n+1}$ 
where $u_n$ is the position of the $n$th C-atom, i.e. the $n$th lattice point). This guarantees the righthand
side of eq.(\ref{ort}) to be dimensionless as it should.
 
The assumption that the hopping parameter $t$ is isotropic and 
constant is correct in 
a first approximation. A spatial dependance of $t$ like
in the SSH model cannot be assumed as phonons are not expected 
to play a significant role 
in the case of flexible chains. A dimerized ground state 
which is vital to the SSH model strictly exists only in one dimension. 
However, the fractal 
dimension of flexible chains in solution 
is $5/3$ and not $1$ \cite{doi:86}. 
 
Nonetheless, the electrons are supposed to move preferably along the chain.
This constraint is imposed by the delta function in eq.(\ref{ort}). 
When "smearing out" this constraint from an interatomic distance $a$ to 
the lower cut-off $l$ of the conformational degrees of freedeom, 
this choice gives the simplest non-trivial coupling between conformation and
electronic degrees of freedom in the continuum limit which will be given 
below.
 
As indicated above, in eq.(\ref{ort}) the electrons are not strictly confined to hop along the
chain. They can also move to and from sites that are next neighbors of the
lattice walk. So one might be tempted to introduce an additional delta
function for the arguments ${\bf x}\pm a{\bf e}_{\mu}$. However, this
procedure would give a wrong continuum limit. It is exactly the "fuzziness" of the electron hopping which
produces the electron-electron interaction that is given below as well as the
stiffening conjectured by Pincus et al.\cite{pincus:87}. In the continuum limit, the apparently superfluous sites fall
on the chain (see eq.(\ref{kontgesamt1})). If the electrons were constrained
to move {\bf only} along the chain on the lattice, then there would be no effect on the
conformation as the sites passed by the electrons could be numbered
systematically. This cannot be done in the case of the interaction given
in eq.(\ref{ort}).  
 
The continuum version of the complete Hamiltonian including the term
for the unperturbed conformation (SAW) is:
\begin{eqnarray}
\label{kontgesamt1}
\beta H &=& 
\frac{d}{2l}\int_0^L ds\left(\pabl{{\bf r}}{s}\right)^2
+\frac{v}{2}\int_0^L ds\int_0^L dt\delta ({\bf r}(s)-{\bf r}(t))\nonumber\\
&&-tl^2\int d^d{\bf x}\left(
c^{\dagger}({\bf x}){\stackrel{\longleftrightarrow}{\nabla^2_{{\bf x}}}}c({\bf x}) 
\right)
\int_0^L ds\delta\left({\bf x}-{\bf r}(s)\right). 
\end{eqnarray}
In the continuum limit it is natural to set $a=l$; $l$ is the Kuhn length that 
represents the minimum characteristic length scale of the system. The present
first model ignores Coulomb interactions between electrons but 
refinements to account for these can be made some of which are mentioned 
below in the context of an effective electron-electron interaction 
(for details see 
\cite{otto:pre94}). 
 
In eq.(\ref{ort}) and (\ref{kontgesamt1}) no preference is given a priori 
to straighter polymer paths. On the contrary, the aim is first, to
choose the SAW for 
the unperturbed polymer conformation and second, to study (perturbatively) 
the effect of
the interaction
term coupling conformation to electronic degrees of freedom. Giving 
preference to stretched conformations would mean a variational treatment of 
the problem which is not necessary here.
 
There are two important aspects to be studied as a consequence of model 
eq.(\ref{kontgesamt1}) in finding the interplay of statistical and electronic
properties. First, the investigation of the effective conformation after eliminating the electronic degrees of freedom. Second, the 
derivation of
an effective interaction between electrons by summing over all 
conformations. 
 
For the following it is also useful to rewrite the Hamiltonian in terms of the collective segment density fields of $N_p$ polymers
$\rho({\bf x})=\sum_{\alpha=1}^{N_p}
\int_0^L ds\delta\left({\bf x}-{\bf r}_{\alpha}(s)\right)$, where 
${\bf r}_{\alpha}(s)$ is 
the position of segment $s$ for chain $\alpha$. The partition function 
to start with is given by:
\begin{equation}
\label{gesamtZ}
Z=\int {\cal D}\rho({\bf x})
\int {\cal D}c^\dagger ({\bf x})\int {\cal D}c({\bf x})
e^{-\beta(H_{conf.}(\{\rho({\bf x})\})+
H_{el}(\{c^\dagger ({\bf x})\},\{c({\bf x})\})},
\end{equation}
where $c^\dagger ({\bf x})$ and $c({\bf x})$ are the Grassmann fields 
for the electrons.
 
For the examination of the effective polymer conformation, an integration over the electron Grassmann fields is performed which yields 
a determinant $\det M$. The result is
an effective Hamiltonian for the conformation including a term 
$-Tr\log M$. For a many polymer system (solutions above the overlap concentrations \cite{degennes:79} or melts) the 
unperturbed conformation may be treated in the Gaussian approximation, i.e. only
terms up to second order in $\rho({\bf x})$ are considered. This will  
not represent a limitation for the validity of the results. To 
handle the determinant the standard
procedures of field theory are applied, i.e. the term $-Tr\log M$ is 
expanded. To remain consistent with the Gaussian approximation the expansion is carried to second order 
in the hopping parameter $t$. The term linear in $t$ gives an overall 
renormalization factor and may be dropped. Finally, the contribution 
to the effective Hamiltonian 
due to the hopping electrons is found to be
\begin{equation}
\label{3.3.13}
\beta H_1 =\frac{1}{V}
\sum_{{\bf k}}
\rho_{-{\bf k}}\rho_{{\bf k}}t^2l^4V^2
\left({\bf k}^4\frac{\Lambda^d}{2d}A+
{\bf k}^2\frac{2\Lambda^{d+2}}{d+2}B
+\frac{2\Lambda^{d+4}}{d+4}A\right),
\end{equation}
where $A=\frac{\Omega_d}{(2\pi)^d}$ and $B=\frac{1}{(2\pi)^d}
\int d\hat{k}\left(1+\cos^2\theta\right)$. $\Lambda$ is a cutoff resulting 
from a ${\bf k}$-space integration and is
the inverse of the polymer cut off (Kuhn length) $l$. This results supports 
the scaling statements of \cite{pincus:87} that the chain shows increased 
stiffness 
due to electron hopping which yields a non-vanishing contribution in 
${\bf k}^4$. The advantage of the present theory is that 
more results beyond the scaling limit can be predicted.
 
To be specific let us look at a concrete model for the unperturbed 
conformation and 
discuss the influence of the hopping parameter $t$ on the parameters of the
conformational model. This is done by examining the structure factor 
$S({\bf k})$. Adopting for a moment the ${\bf k}$-expansion by Shimada et al. 
\cite{shimada:88a} which discusses stiff chains in the limit of moderate bending energies, the
contribution of the electrons enter the first coefficients as follows:
\begin{eqnarray}
\label{3.3.17}
S^{-1}({\bf k}) &=&\frac{1}{\rho L}[1+\rho L(v+t^2\tilde{a})
+{\bf k}^2(\frac{1}{9}Ll_p+t^2\tilde{b}\rho L)\nonumber\\
&&+{\bf k}^4(\frac{Ll_p}{324}+t^2\tilde{c}\rho L)+...].
\end{eqnarray}
where $\tilde{a}=\frac{2\Lambda^{d+4}}{d+4}Aa^4V^2$, 
$\tilde{b}=\frac{2\Lambda^{d+2}}{d+2}Ba^4V^2$, and
$\tilde{c}=\frac{\Lambda^{d}}{2d}Aa^4V^2$.
We thus find - for the single chain contour hopping model eq.(\ref{ort}), 
neglecting other intra and
inter chain hopping processes - contributions to the excluded volume effect
as well as to terms in second and fourth power of ${\bf k}$. Especially 
the two latter contributions have to be interpreted in terms of an 
increased persistence
length as the hopping parameter is tuned up.
 
The second aspect of the model, the influence of the conformation on the
interaction between the electrons, becomes manifest if the density variables 
in eq.(\ref{gesamtZ}) are integrated out. Then in order to obtain 
the effective Hamiltonian for the
electrons a cumulant expansion may be carried out in momentum space. This
expansion can be visualized in terms of Feynman graphs where the structure 
factor takes the role of a propagator for a "particle exchanged" between
electrons.
In second order of $t$ the Fourier transform of the
effective inter electron potential is extracted. It is proportional to the 
conformational structure factor. Different models for the unperturbed 
conformation can be studied. For simplicity here only Gaussian chains (without
excluded volume) are investigated. For such random walk chains the 
pair potential in $d$ spatial dimensions as a function of the scaled 
distance $x=r\sqrt{\frac{3d}{2Ll}}$ is:
\beq
\label{potort}
\beta V(x)=-\frac{2}{(2\pi)^{\frac{d}{2}}}t_0^2\beta^2
\left(\frac{3d}{lL}\right)^{\frac{d}{2}+2}N_pVl^4
\left(\frac{d}{2}(\frac{d}{2}+1)-(d+2)x^2+x^4\right)
e^{-x^2}.
\eeq
Its behavior is shown in fig.(\ref{drawpot}). A magnified view of the 
non-zero minimum is included. The prefactors in eq.(\ref{potort})
have been set equal to one to give the indicated scale.
As fig.(\ref{drawpot}) demonstrates, the pair potential has an attractive 
short-range part, a repulsive part for intermediate length scales which gradually
decreases to zero for long distances. The attractive short-range part 
points to a disorder-induced localization of the $\pi$-electrons due 
to conformational entropy. Coulomb correlations are not included in 
eq.(\ref{potort}). If they are treated as in the classical Hubbard model, they contribute 
a positive singularity at the origin (in the form of Dirac's delta function).
At room temperatures $\beta\Delta V\sim 10^{-2}$, so electrons are likely to 
overcome the attractive part of the potential in this case. The attractive 
part of the potential becomes important forcing electrons to localize at 
temperatures where the flexibility of a real polymer is completely lost.
 
The analysis of effective electron-electron interactions may be carried 
further by including more complicated hopping processes. They lead to 
seagull-type and other Feynman graphs which will be discussed elsewhere 
\cite{otto:pre94}. These depend on higher (3-point and 4-point) correlation function of
the density variables. In terms of conformational interactions these correspond to
the inclusion of the second and third virial coefficient. Thus the series 
remains finite 
if these coefficients vanish which they actually do in the case of Gaussian 
chains.
A proper $O(n)$-field theory (in the limit $n\rightarrow 0$ )
for such conducting chains can be easily
constructed, if a SAW \cite{degennes:72} is used for the underlying conformation of the non-conducting chain. The $n$ component vector fields 
${\bf \phi}$ 
are coupled to fermion fields for the hopping 
electrons \cite{otto:zph94}. An effective field theory for the conformation can be obtained 
by integrating over the electron fields. This gives rise, among other effects,
to a renormalized excluded volume parameter:
\begin{equation}
\label{3.4.32}
v_{1}=v+const\frac{1}{d+4}\frac{\beta^2 t_0^2}{L^2}V^2l^{-d}.
\end{equation}
$v$ is the excluded volume parameter of the non-conducting SAW.
One also obtains higher gradient terms in the conformation fields that can 
be eliminated by counter terms. A supersymmetric field theory following 
\cite{parisi:80}\cite{efetov:83} may also be conceived. Indeed such a theory 
confirms the results found in the bosonic version together with the limit $n\rightarrow 0$.
 
How does $v_1$ affect the conformation? A simple estimate is given by 
a Flory-type argument that starts from a crude approximation to the free 
energy for the modified SAW:
\begin{equation}
\label{skalen.1}
\beta F=\frac{dR^2}{2Ll}+\frac{v_1}{2}\frac{L^2}{R^d}.
\end{equation}
Now $v_1$ has to be implemented according to eq.(\ref{3.4.32}). The volume 
$V$ is given by the mean volume of the chain, $R^d$. If $R$ is expressed in 
terms of the contour length $L$, one finds in the case of Gaussian chains 
$V\sim L^{\frac{d}{2}}$. The excluded volume parameter $v_1$ is therefore 
given by $v_1=v_0+C\beta^2t_0^2L^{d-2}$. Substituting this expression for $v_1$
in eq.(\ref{skalen.1}) and minimizing with respect to $R$ yields the mean 
end-to-end vector. For $\beta^2t_0^2L^{d-2}\gg v_0$ we find $R\propto L^{\nu}$ 
with a new Flory exponent:
\begin{equation}
\label{skalen.7}
\nu_{Flory,Hopping}=\frac{d+1}{d+2}.
\end{equation}
In $d=3$ it is $0.8$ whereas the common exponent for SAW reads roughly as $0.6$. This 
implies a strong swelling and stretching of the chain. Although the term
induced by the hopping of the electrons dominates over the excluded volume we
do not find a completely stretched chain, i.e. we cannot find a rod-to-coil
transition. It has to be pointed out also that this term yields no finite upper critical dimension. The 
hopping 
term is important in all space dimensions, which is in accord with the 
physical intuition. The hopping of the electron is purely local from 
segment to segment and is not sensitive to the space dimension, whereas the usual 
excluded volume interaction becomes irrelevant in $d>4$. 
 
We have presented a field theory for conducting, flexible self-avoiding walks. Starting from the path integral formulation of the chain Hamiltonian and a 
Hubbard model where the electron hopping is bound to take place on the chain, several results could be found. 
First the hypothesis of Pincus 
et al. could be confirmed that stiff or rod-like conformations 
are favored due to a strong interplay of entropic conformation of the polymers and the electronic energy. Moreover a new exponent for the size of a conducting SAW could be derived. The chains are significantly stretched. However, even 
in the Flory approximation no rods are formed. This is in contrast to
Pincus et al. as far as the explicit assumption of extended rod-like segments
made in \cite{pincus:87} is concerned. The result found above agrees with the physical picture that the $\pi$-electrons are likely to 
delocalize more easily when the chains are stretched. Therefore they are able to contribute to the electronic properties. 
One advantage of the field theoretic formulation is that the influence of the conformation on the effective pair potential between the electrons can be calculated. 
It shows a repulsive and an attractive part, 
which produces a second minimum. Several extensions and modifications of this theory are possible and will be reported separately.
%
%

%
\newpage
\begin{figure}[h]
  \caption[]{The pair potential in $d=3$; the inset
gives a magnified view of the second minimum.}
  \label{drawpot}
\end{figure}
\newpage
\begin{figure}[htpb]
  \begin{center}
  \begin{minipage}{15cm}
  {\epsfxsize 15 cm
\epsfbox{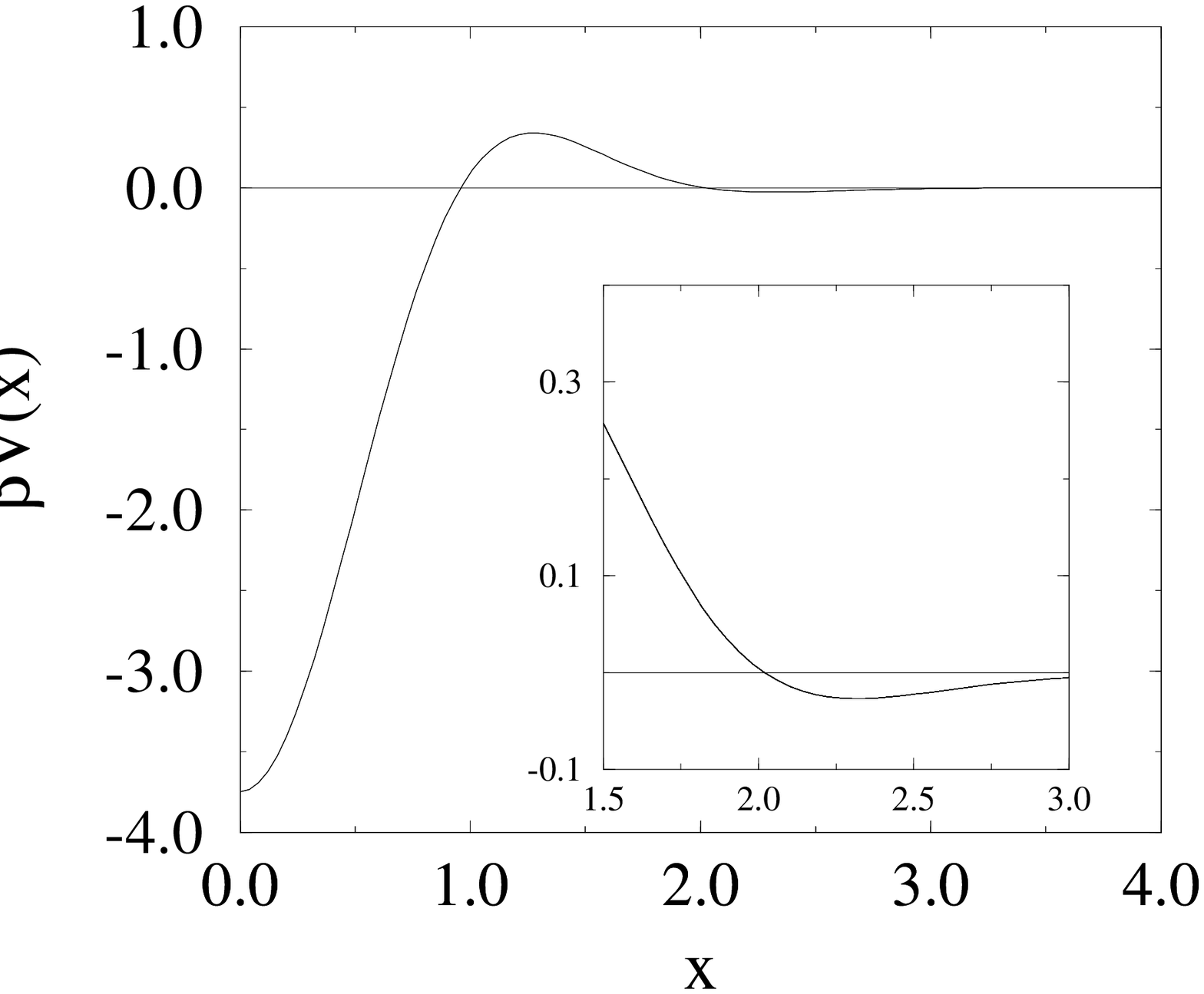}}
  \end{minipage}
  \end{center}
\end{figure}
\end{document}